\documentstyle[12pt,epsf,aasms4]{article}

\def\gtorder{\mathrel{\raise.3ex\hbox{$>$}\mkern-14mu
                \lower0.6ex\hbox{$\sim$}}}
\def\ltorder{\mathrel{\raise.3ex\hbox{$<$}\mkern-14mu
                \lower0.6ex\hbox{$\sim$}}}
\newcommand{\hii}{H~{\small II} }

\begin{document}

\title{A New Catalog of Radio Compact \hii Regions in the Milky Way}

\author{Giveon U. (UC Davis), Becker R.H. (UC Davis \& LLNL), Helfand, D.J. (Columbia University), and White R.L. (STScI)}

\begin{abstract}
We utilize new VLA Galactic plane catalogs at 5 and 1.4 GHz covering the first Galactic quadrant ($350^o\le l\le 42^o$, $|b|\le 0.4^o$) in conjunction with the MSX6C Galactic plane catalog to construct a large sample of ultra-compact \hii regions. A radio catalog of this region was first published by Becker et al. (1994), but we have added new observations and re-reduced the data with significantly improved calibration and mosaicing procedures, resulting in a
tripling of the number of 5 GHz sources detected. Comparison of the new 5 GHz catalog and the MSX6C Galactic plane catalog resulted in a sample of 687 matches, out of which we estimate only 15 to be chance coincidences. Most of the matches show red MSX colors and a thermal radio spectrum. The scale height of their Galactic latitude distribution is very small (FWHM of $16'$ or $\sim 40$ pc). These properties suggest that the sample is dominated by young ultra-compact \hii regions, most of which are previously uncataloged.
\end{abstract}

\section{Introduction}
\label{intro}

Both young and old stars are often embedded in clouds of dust and cold gas, obscuring them at wavelengths shortward of the near-infrared range. This includes star-forming regions, in which new stars and star clusters are born in huge molecular clouds, as well as various classes of evolved stars: e.g., planetary nebulae (PNe) and cool giant stars. Some of these sources are hot enough to ionize the gas in their vicinity, producing radio emission via free-free electron interactions. Cooler molecular gas and dust which still shroud these ionized regions absorb the radiation from the central source and reradiate it in the infrared. As has been shown previously, combining infrared and radio surveys can be a powerful tool in identifying specific classes of objects (i.e., \hii regions, ultra-compact [UC] \hii regions, and planetary nebulae) based on the simultaneous presence of the cold molecular and dust phases and the hot ionized gas (White, Becker \& Helfand 1991; Becker et al. 1994; Szymczak et al. 2002).

Earlier studies have employed different tracers of infrared and radio emission at various levels of sensitivity and resolution according to the targets of interest.
Most authors have used photometric measurements from IRAS or ISO to obtain infrared colors (e.g., Becker et al. 1994; Mart\'{i}n-Hern\'{a}ndez, van der Hulst, \& Tielens 2003). Observations from the Midcourse Space Experiment (MSX) satellite mission (Price et al. 2001; Egan, Price \& Kramer 2003) have rarely been used in combination with radio observations (see, however, work by Cohen \& Green 2001 and by Kerton 2002).
In the radio range, a variety of measures are used depending on the type of sources being investigated and the information needed. Radio emission lines give information on the velocity and distance of the source (Mooney 1992; Kurtz et al. 1994; Hayashi et al. 1994; Bourke, Hyland, \& Robinson 1995; Walsh et al. 1995; Carpenter, Heyer, \& Snell 2000; Sevenster 2002; Yang et al. 2002 ; Kerton \& Brunt 2003 ), while continuum measurements are used to estimate the source ionizing flux (White, Becker \& Helfand 1991; Becker et al. 1994; Szymczak et al. 2002). A number of extragalactic studies have used the technique of matching infrared and radio catalogs to isolate and study specific populations such as dust enshrouded star-forming galaxies and active galactic nuclei (Condon et al. 2002; Hopkins et al. 2003).

In this work we utilize new VLA Galactic plane catalogs at 5 and 1.4 GHz, the original versions of which were published by Becker et al. (1994) and Zoonematkermani et al. (1990), respectively. We compare the 5 GHz catalog to the most recent version from the MSX.
In \S\ref{6cm} the new radio catalogs are briefly introduced.
In \S\ref{iradio} we describe the matching of the 5 GHz radio catalog with the MSX6C catalog, and in \S\ref{properties} we study the spatial and color properties of the resulting subset of sources. In \S\ref{conc} we summarize our conclusions.

\section{The VLA Galactic Plane Survey at 5 GHz}
\label{6cm}

The original 5 GHz observations, described in detail in Becker et al. (1994), covered the inner Galaxy ($350^o\le l\le 42^o$, $|b|\le 0.4^o$) with a resolution of $\sim 4''$ and sensitivity of $\sim 2.5$ mJy. Most of the original observations were taken in the VLA C configuration, but some fields were observed in the CnD configuration. The CnD fields were re-observed in early 2004 in C array to create a more uniform survey. Incorporating these new data, we applied improved phase calibration and reduction procedures, resulting in an increase in the number of sources detected by nearly a factor of three from 1272 entries to 
3283.
The principal improvement resulted from mosaicing areas of overlapping coverage which improved map sensitivity, particularly at the edges of the VLA primary beam. The new 5 GHz catalog, to be published separately (Becker, White, \& Helfand 2004), serves as the primary radio catalog for our comparison with the MSX6C catalog.

The new improved observations and reduction increased significantly the effective sensitivity of the 5 GHz catalog; it is now $>90\%$ complete for sources with $F_{5 GHz}\ge 3$ mJy, compared to 85\% completeness for 10 mJy sources in the earlier version. Adopting stellar fluxes from Sternberg, Hoffmann, \& Pauldrach (2003), we calculate that this allows us to detect at the far edge of the Galaxy (distance of 20 kpc) a B0 star with a nebula that is optically thin to free-free emission ($F_{5 GHz}=30$ mJy), or an O9 star with an optically thick nebula ($F_{5 GHz}=3$ mJy for an emission measure of $\sim 10^9$ pc$\cdot$cm$^{-6}$). Expected radio and infrared flux densities for the hottest main sequence stars are listed in Table \ref{spec_type}. Infrared flux densities were calculated assuming 1\% of the total luminosity is reradiated in the MSX 8 $\mu$m band (based on the IRAS 12 $\mu$m band response stated by Wood \& Churchwell [1989] and on the MSX 8 $\mu$m band response function). The emission measure we used for the optically thick case corresponds in the most extreme case to a nebular gas density of $10^6$ cm$^{-3}$ and a nebular diameter 0.003 pc (e.g., Mart\'{i}n-Hern\'{a}ndez, van der Hulst, \& Tielens 2003; Olmi \& Cesaroni 1999). Based on this table and on the sensitivity of the survey, we conclude that the radio survey is $>90\%$ complete in detecting all O stars across the Galactic disk in the region observed, although in \S \ref{iradio} we show that the match to the MSX6C catalog significantly reduces this completeness level.
\begin{table}
\begin{tabular}[h]{ccccccc} \hline
Spectral & $T_{eff}$ & $\log$ L      & $\log$ Q      & $F_{8\mu m}$ & $F_{5GHz}$ ($\tau\ll 1$) & $F_{5GHz}$ ($\tau\gg 1$) \\
Type     & [K]       & [$L_{\odot}$] & [$\#/s^{-1}$] & [Jy]     & [mJy]               & [mJy]               \\
\hline\hline
O3       &     51000  &   6.3       &  49.9         & 2.2        &   2100      &   84        \\
O5       &     46000  &   6.0       &  49.5         & 1.1       &    840      &   32        \\
O7       &     41000  &   5.7       &  49.1         & 0.55       &    330      &   13        \\
O9       &     36000  &   5.4       &  48.5         & 0.28      &     84      &    3.3      \\
O9.5     &     35000  &   5.3       &  48.3         & 0.22      &     53      &    2.1      \\
B0       &     33000  &   5.2       &  48.0         & 0.17      &     26      &    1.0      \\
B0.5     &     32000  &   5.1       &  47.7         & 0.14      &     13      &    0.5      \\
\hline
\end{tabular}
\caption{Radio and infrared flux densities, for a distance of 20 kpc, based on model stellar spectra (Sternberg, Hoffmann, \& Pauldrach 2003). Columns list spectral types, effective temperatures, bolometric luminosities, ionizing photon fluxes $Q$, 8 $\mu$m flux densities, and 5 GHz flux densities for optically thin and optically thick nebulae. We have calculated the infrared flux densities assuming 1\% of the total luminosity is reradiated in the MSX 8 $\mu$m band (based on the IRAS 12 $\mu$m band response stated by Wood \& Churchwell [1989] and on the MSX 8 $\mu$m band response function). Optically thick 5 GHz flux densities were calculated using an emission measure of $\sim 10^9$ pc$\cdot$cm$^{-6}$.}
\label{spec_type}
\end{table}

\section{Radio-Infrared Matching}
\label{iradio}

As the combination of radio and infrared emission can be employed to distinguish and investigate populations of Galactic sources, we perform a matching analysis based on positions between the new 5 GHz catalog and the MSX6C infrared catalog.
The MSX6C archival data (version 2.3, Egan, Price, \& Kramer 2003) comprises 
431 711 sources observed in four bands -- A, C, D, and E  with central wavelengths 8.3, 12.1, 14.7, and 21.3 $\mu$m, respectively. Hereafter we will refer to the MSX bands by their central wavelengths -- 8, 12, 14, and 21 $\mu$m. The MSX6C survey covers the entire Galaxy for latitudes $|b|\le 6^o$ with a resolution of $18.3''$. The sensitivity ranges from 0.1 Jy at 8 $\mu$m to 6 Jy at 21 $\mu$m. The MSX6C coverage is not uniform over this area owing to varying scan overlap resulting from a change in the scan rate at the end of the mission (Egan, Price, \& Kramer 2003).

We match the 5 GHz catalog and the MSX6C catalog by considering two criteria: $\alpha$ - a conservative matching radius of $12''$, and $\beta$ - an additional annulus with $12''<r\le 25''$; adding this larger annulus still yields real matches. We estimate the number of false matches by comparing the MSX6C catalog to false radio catalogs with the same spatial distribution by shifting the original catalog by $\pm 10'$ and $\pm 20'$ in Galactic longitude. The results from the matches for each criterion are given in Table \ref{proba1}: number of total matches (column 3), and number of false matches (column 4) from the shifted radio maps. In the real catalog, however, the false match rate will be lower than in the shifted catalog since true matches cannot also be false matches. Hence, we correct the false match rate for the fraction of true matches by a factor of $1-F/N$, where $F$ is the number of false matches and $N$ is the total number of sources in the radio catalog. The true match rates that correspond to these corrected false rates are listed in column 5.

\begin{table}
\caption{Matching criteria and results.}
\begin{tabular}[h]{ccccc} \hline
Matching  & Matching & Total   & False   & \% True \\
Criterion & Radius   & Matches & Matches & Matches \\ 
(1) & (2) & (3) & (4) & (5) \\ \hline\hline
$\alpha$ &$12''$     & 804 & 131 & 87 \\
$\beta$  &$12''-25''$& 650 & 366 & 49 \\
\hline
\end{tabular}
\label{proba1}
\end{table}

We have used the information on the MSX6C bands in which a given source was detected to refine the probability calculation in Table \ref{proba1} (White, Becker \& Helfand 1991). In principle, the distribution of false matches should mimic the MSX6C band distribution, while MSX6C sources with 5 GHz emission may have quite a different distribution of band detections. Table \ref{proba2} lists the results of these probability calculations. We have counted the number of detections in each band combination for the MSX6C sources in the Galactic region roughly defined by the radio catalog ($348^o\le l\le 42^o$, $|b|\le 0.5^o$). Column 2 lists these numbers, while column 3 lists their percentage of the total number of sources. We have calculated the expected false rates (columns 4 \& 5) for these numbers according to the values given in Table \ref{proba1}.
The same correction needs to be applied to the multi-band match. Here the correction factor to be applied comes from the sum of true and false matches in all bands. Let $M_i$ be the actual number of matches to band $i$, $F_i$ be the number of false matches, and $N$ be the total number of sources in the radio catalog. The corrected formula for the reliability is
\begin{equation}
R_i=1-{{F_i}\over{M_i}}\cdot{{N-\sum M_j}\over{N-\sum F_j}}
\label{match_eq}
\end{equation}
where the sums are over all the bands, j=1..15. Effectively, the false rate in each band is reduced by the factor $(N-\sum M_j)/(N-\sum F_j)$. If $\sum M_j=\sum F_j$, the factor is 1 and all the matches are false. If $F_i=0$, the factor is 0 and all the matches are real. The sum of this equation over all the bands gives the same overall estimate of the match rates in Table \ref{proba1}.
The actual matches found in each band combination are given in columns 6 \& 7, while the final $R_i$ values are given in columns 8 \& 9.

The physical interpretation is that the false match experiments tell how many unmatched objects should be left over from the false matches. If the number left over from the true match is smaller than that, there must have been fewer unmatched objects to begin with (because some had true matches), so the expected number of false matches is smaller by the ratio of unmatched objects.
\begin{table}
\scriptsize
\begin{tabular}[h]{lcccccccc} \hline
MSX    & Number of & \% of    & \multicolumn{2}{c}{Expected False Matches} & \multicolumn{2}{c}{Actual Total Matches} & \multicolumn{2}{c}{Reliability} \\
Bands & MSX Sources$^a$    & MSX Sources & $\alpha$ & $\beta$ & $\alpha$     & $\beta$ & $\alpha$ & $\beta$ \\ \hline
(1) & (2) & (3) & (4) & (5) & (6) & (7) & (8) & (9) \\ \hline\hline
8 only$^b$     & 29 274 & 51.78 & 68    & 190    &  46 & 126 & 0.00     & 0.00     \\
12 only    &      0 &  0    &  0    &   0    &   0 &   0 & $\cdots$ & $\cdots$ \\
14 only    &      0 &  0    &  0    &   0    &   0 &   0 & $\cdots$ & $\cdots$ \\
21 only    &     37 &  0.07 &  0.09 &   0.25 &   0 &   0 & $\cdots$ & $\cdots$ \\
8-12       &   5061 &  8.95 & 12    &  33    &  17 &  41 & 0.45     & 0.27     \\
8-14       &   3216 &  5.69 &  7.5  &  21    &  21 &  23 & 0.72     & 0.18     \\
8-21       &   1248 &  2.21 &  2.9  &   8.1  &  11 &   8 & 0.79     & 0.09     \\
12-14      &      1 &  0    &  0    &   0    &   0 &   0 & $\cdots$ & $\cdots$ \\
12-21      &      3 &  0    &  0    &   0    &   0 &   0 & $\cdots$ & $\cdots$ \\
14-21      &     98 &  0.17 &  0.2  &   0.6  &   3 &   2 & 0.95     & 0.73     \\
8-12-14    &   8086 & 14.30 & 19    &  52    &  22 &  48 & 0.32     & 0.02     \\
8-12-21    &    529 &  0.94 &  1.2  &   3.4  &  14 &  11 & 0.93     & 0.72     \\
8-14-21    &    497 &  0.88 &  1.2  &   3.2  &  25 &  17 & 0.96     & 0.83     \\
12-14-21   &    130 &  0.23 &  0.3  &   0.8  &  12 &  15 & 0.98     & 0.95     \\
8-12-14-21 &   8348 & 14.78 & 19    &  54    & 633 & 359 & 0.98     & 0.86     \\
\hline
\end{tabular}
\caption{Matching reliability using band detection information.\newline
$^a$ Taken from a limited area with coordinates ranges approximately the same as the 5 GHz catalog: $348^o\le l\le 42^o$, $|b|\le 0.5^o$.\newline
$^b$ The expected false rate in this band combination is significantly overestimated. The large discrepancy between expected and actual false rates is a result of these sources being mainly stars, which have a flat latitude distribution in the survey area, while the distribution of the radio sources peaks at the Galactic plane.}
\label{proba2}
\end{table}
We find that 89\% of the $\alpha$ criterion matches are $\ge50\%$ reliable, and 62\% of the $\beta$ criterion matches are $\ge 50\%$ reliable. A total of 702 matches from both criteria have reliability $\ge 90\%$ of which 16 are expected to be false. Of these, 645 are reliable at $\ge 98\%$. This is a major improvement over the color-independent values in Table \ref{proba1}. For point sources (radio size $\le 5''$) with high-reliability matches ($\ge 0.98$) we find that the two catalogs' astrometry is consistent within errors of $3''$.

Out of the 702 matches with 90\% reliability, 14 are multiple MSX6C matches -- cases where one radio source has two or three MSX6C counterparts within the matching radius -- and 141 are multiple radio matches -- cases where one MSX6C source has two or more radio counterparts. For the latter case, sometimes one of the radio matches has significantly lower reliability, and it is possible that it is a false match, but in most cases the reliabilities are comparable and the multiplicity implies real clustering. This clustering could also result from the coarser spatial resolution of the MSX6C catalog. The 14 multiple MSX6C matches correspond to 29 matching entries, so for our subsequent analysis we exclude the 15 additional entries to get 687 ($=702-15$) single radio sources that have high reliability ($\ge 90\%$) MSX6C matches, where 15 of them are estimated to be false matches. The total number of single sources with reliability $\ge 50\%$ is 1074, with 60 estimated to be false matches.
Table \ref{source_list} lists the first 50 sources from this catalog; the full tables of all high-reliability sources and of the additional 387 low-reliability sources are available electronically. For each source we provide the Galactic coordinates, 5 GHz peak and integrated flux densities in mJy, deconvolved major and minor axes in arcseconds, distance from the MSX6C source in arcseconds, name of the MSX6C match, 8, 12, 14, and 21 $\mu$m flux densities of the MSX6C match in Jy, and the reliability of the match. 
\begin{table}
\tiny
\begin{tabular}[h]{crrrrrcrrrrc} \hline
Name & $F_p$ & $F_i$ & Major  & Minor  & Separation & MSX6C Name & $F_8$ & $F_{12}$ & $F_{14}$ & $F_{21}$ & Reliability \\
     & (mJy) & (mJy) &(arcsec)&(arcsec)&  (arcsec)  &            &  (Jy) &  (Jy)    &   (Jy)   &   (Jy)   &             \\ \hline
(1) & (2) & (3) & (4) & (5) & (6) & (7) & (8) & (9) & (10) & (11) & (12) \\ \hline\hline
350.09340+0.23145   &   71.1 &   73.5 &  1.43 &  0.00 &  2.8 & G350.0922+00.2306 &    0.91 &    1.62 &    5.39 &   18.85 & 0.98 \\
350.09319+0.08965   &   10.6 &   15.5 &  3.98 &  3.48 &  5.2 & G350.0927+00.0888 &    2.70 &    9.86 &   11.56 &   31.72 & 0.98 \\
350.10463+0.08153   &  235.9 &  611.5 &  7.99 &  5.87 & 11.8 & G350.1051+00.0795 &    8.16 &   21.43 &   26.44 &   95.57 & 0.98 \\
350.10740+0.07921   &   31.1 &   78.8 &  9.79 &  5.57 &  4.3 & G350.1051+00.0795 &    8.16 &   21.43 &   26.44 &   95.57 & 0.98 \\
350.11934+0.06978   &   95.2 &  154.7 &  7.04 &  3.19 &  2.7 & G350.1184+00.0692 &    5.31 &   14.71 &   19.83 &   66.47 & 0.98 \\
350.17375+0.02880   &    7.9 &   12.8 &  5.29 &  3.28 &  4.6 & G350.1723+00.0274 &    0.73 &    1.03 &    0.99 &    3.94 & 0.98 \\
350.34187+0.14172   &    9.2 &    9.2 &  1.08 &  0.00 &  9.9 & G350.3395+00.1435 &    0.80 &    1.19 &    0.92 &    3.72 & 0.98 \\
350.24594+0.06493   &    4.2 &    8.4 &  7.41 &  4.13 &  2.3 & G350.2443+00.0641 &    1.95 &    2.52 &    2.40 &    5.76 & 0.98 \\
350.18148+0.01475   &  107.3 &  266.5 &  7.13 &  6.56 &  6.7 & G350.1783+00.0147 &    1.16 &    2.01 &    1.93 &    7.83 & 0.98 \\
350.33215+0.09986   &   50.8 &   63.4 &  3.03 &  1.99 &  2.3 & G350.3305+00.0990 &    1.73 &    5.63 &    9.15 &   36.24 & 0.98 \\
350.28232$-$0.03533 &    8.2 &    9.9 &  4.56 &  0.00 &  5.2 & G350.2816-00.0363 &    0.40 &    0.79 &    0.76 &    3.00 & 0.98 \\
350.78265$-$0.02745 &   98.9 &  152.0 &  6.88 &  1.76 &  8.4 & G350.7833-00.0273 &    6.47 &   24.75 &   45.35 &  113.51 & 0.98 \\
350.78416$-$0.02754 &  624.3 & 1074.6 &  5.71 &  3.39 &  3.0 & G350.7833-00.0273 &    6.47 &   24.75 &   45.35 &  113.51 & 0.98 \\
350.77438$-$0.07850 &    4.4 &   15.9 & 20.84 &  3.61 &  1.2 & G350.7728-00.0790 &    1.00 &    0.96 &    0.93 &    6.12 & 0.98 \\
350.95351$-$0.08396 &   15.9 &   21.4 &  5.88 &  0.00 &  3.2 & G350.9528-00.0841 &    0.55 &    1.28 &    2.22 &    5.52 & 0.98 \\
351.55662+0.20500   &  637.4 & 1790.6 &  9.05 &  6.79 &  5.2 & G351.5556+00.2037 &   24.22 &   93.18 &  138.18 &  313.59 & 0.98 \\
351.61288+0.17168   &   51.9 &  254.2 & 22.27 &  6.10 &  6.3 & G351.6126+00.1725 &    5.09 &   30.83 &   38.78 &  113.71 & 0.98 \\
351.61227+0.16619   &   77.2 &  449.8 & 17.60 &  9.43 &  6.0 & G351.6120+00.1652 &   21.24 &   96.37 &  154.87 &  556.77 & 0.98 \\
351.61400+0.16559   &  317.0 &  707.3 &  6.47 &  5.53 &  1.9 & G351.6120+00.1652 &   21.24 &   96.37 &  154.87 &  556.77 & 0.98 \\
351.04265$-$0.33388 &  111.2 &  148.3 &  3.35 &  2.89 &  7.0 & G351.0429-00.3339 &    2.86 &    2.46 &    2.25 &    7.77 & 0.98 \\
352.08661+0.15174   &    6.5 &   14.5 &  6.59 &  6.02 &  4.6 & G352.0860+00.1509 &    0.77 &    1.17 &    1.11 &    3.31 & 0.98 \\
352.10488+0.16202   &   11.1 &   36.9 &  8.92 &  8.34 &  1.5 & G352.1033+00.1621 &    0.96 &    1.34 &    1.27 &    3.42 & 0.98 \\
351.58721$-$0.35006 &   66.9 &  144.1 &  7.66 &  4.48 &  7.5 & G351.5876-00.3501 &    0.09 & $<$0.65 &    0.94 &    6.00 & 0.96 \\
352.61685+0.17544   &    4.2 &   18.0 & 18.66 &  5.99 & 11.7 & G352.6122+00.1740 & $<$0.04 &    1.58 &    2.27 &    6.54 & 0.98 \\
352.61396+0.17284   &    4.9 &   14.1 & 11.12 &  5.99 &  6.1 & G352.6122+00.1740 & $<$0.04 &    1.58 &    2.27 &    6.54 & 0.98 \\
352.67634+0.14480   &  311.6 &  591.7 &  5.50 &  4.75 &  5.4 & G352.6755+00.1435 &    2.34 &    3.46 &   18.57 &   36.00 & 0.98 \\
352.39355$-$0.06332 &    4.1 &   13.1 & 24.30 &  0.00 & 11.4 & G352.3938-00.0654 &    3.77 &    8.04 &    8.28 &   21.80 & 0.98 \\
352.39534$-$0.06494 &   12.0 &   36.8 & 11.19 &  5.81 &  1.0 & G352.3938-00.0654 &    3.77 &    8.04 &    8.28 &   21.80 & 0.98 \\
352.58866$-$0.18062 &   13.9 &   96.1 & 18.74 &  9.42 & 10.3 & G352.5844-00.1810 &    2.95 &    5.98 &    3.90 &   13.52 & 0.98 \\
352.60087$-$0.17647 &    9.8 &   73.2 & 16.01 & 13.62 &  7.8 & G352.5975-00.1762 &    3.89 &   16.53 &   22.31 &   41.85 & 0.98 \\
352.60824+0.17634   &    2.6 &    8.1 & 10.34 &  4.37 & 24.8 & G352.6122+00.1740 & $<$0.04 &    1.58 &    2.27 &    6.54 & 0.95 \\
352.61038+0.17559   &    1.6 &    1.7 &  3.20 &  0.00 & 15.4 & G352.6122+00.1740 & $<$0.04 &    1.58 &    2.27 &    6.54 & 0.95 \\
352.61827+0.17920   &    2.5 &    4.3 &  6.32 &  4.08 & 24.2 & G352.6122+00.1740 & $<$0.04 &    1.58 &    2.27 &    6.54 & 0.95 \\
353.45400+0.32568   &   11.0 &   42.9 & 10.03 &  9.59 &  3.0 & G353.4523+00.3246 &    0.56 &    0.79 &    0.91 &    1.74 & 0.98 \\
352.85820$-$0.20367 &   51.7 &  197.4 & 11.81 &  7.77 &  4.2 & G352.8577-00.2040 &    7.25 &   19.09 &   22.36 &   53.01 & 0.98 \\
352.85891$-$0.20596 &   11.2 &   49.0 & 17.82 &  0.00 &  9.0 & G352.8577-00.2040 &    7.25 &   19.09 &   22.36 &   53.01 & 0.98 \\
352.83039$-$0.25758 &  154.1 &  458.3 &  9.03 &  7.37 &  6.8 & G352.8282-00.2594 &    3.19 &    2.45 &   20.38 &   33.97 & 0.98 \\
353.39799$-$0.06923 &   16.0 &   20.6 &  4.68 &  1.58 &  0.4 & G353.3965-00.0694 &    0.60 &    0.93 &    1.02 &    6.71 & 0.98 \\
353.39657$-$0.09544 &   12.2 &   42.2 & 11.86 &  7.34 &  4.9 & G353.3960-00.0963 &    1.24 &    2.13 &    3.08 &    7.10 & 0.98 \\
353.54841$-$0.01341 &   16.3 &  127.5 & 17.30 & 13.31 & 11.7 & G353.5457-00.0164 &    6.98 &   19.98 &   26.20 &   72.24 & 0.98 \\
353.54733$-$0.01583 &   98.1 &  210.8 &  7.14 &  2.98 &  1.6 & G353.5457-00.0164 &    6.98 &   19.98 &   26.20 &   72.24 & 0.98 \\
353.54787$-$0.01887 &   25.2 &  184.6 & 20.96 &  9.72 & 11.6 & G353.5457-00.0164 &    6.98 &   19.98 &   26.20 &   72.24 & 0.98 \\
353.36439$-$0.16474 &  248.7 &  314.1 &  2.92 &  2.08 &  7.0 & G353.3633-00.1665 &    5.99 &   10.45 &   15.13 &   73.43 & 0.98 \\
354.04635+0.22847   &    6.6 &    7.3 &  4.35 &  0.00 &  0.7 & G354.0446+00.2282 &   21.06 &   19.15 &   16.74 &    9.59 & 0.98 \\
353.42191$-$0.36904 &   60.1 &  235.4 & 19.30 &  4.34 &  8.3 & G353.4213-00.3709 &   10.44 &   23.74 &   21.32 &   76.41 & 0.98 \\
353.42136$-$0.37245 &   52.7 &   75.3 &  7.51 &  0.00 &  8.5 & G353.4213-00.3709 &   10.44 &   23.74 &   21.32 &   76.41 & 0.98 \\
353.79394$-$0.15955 &    8.6 &   18.4 &  6.22 &  5.41 &  1.6 & G353.7928-00.1600 &    1.53 &    2.17 &    1.35 &    2.59 & 0.98 \\
354.72595+0.30053   &  189.5 &  253.5 &  3.72 &  2.64 &  1.4 & G354.7241+00.3001 &    3.06 &    8.97 &   14.88 &   42.25 & 0.98 \\
354.19028$-$0.05879 &   49.7 &  231.9 & 11.92 &  9.86 &  3.5 & G354.1882-00.0598 &    6.76 &   16.92 &   21.23 &   57.76 & 0.98 \\
354.71436+0.28882   &   23.6 &   90.4 & 11.61 &  7.94 &  5.1 & G354.7130+00.2873 &    0.58 &    0.63 &    0.76 &    2.97 & 0.98 \\
\hline
\end{tabular}
\caption{First 50 entries of high-reliability 5 GHz-MSX6C matches.}
\label{source_list}
\end{table}

\section{Properties of the Matching Subset}
\label{properties}

\subsection{Infrared Colors}
\label{rad_ir}

\begin{figure}
\centerline{\epsfxsize=7.0in\epsfbox{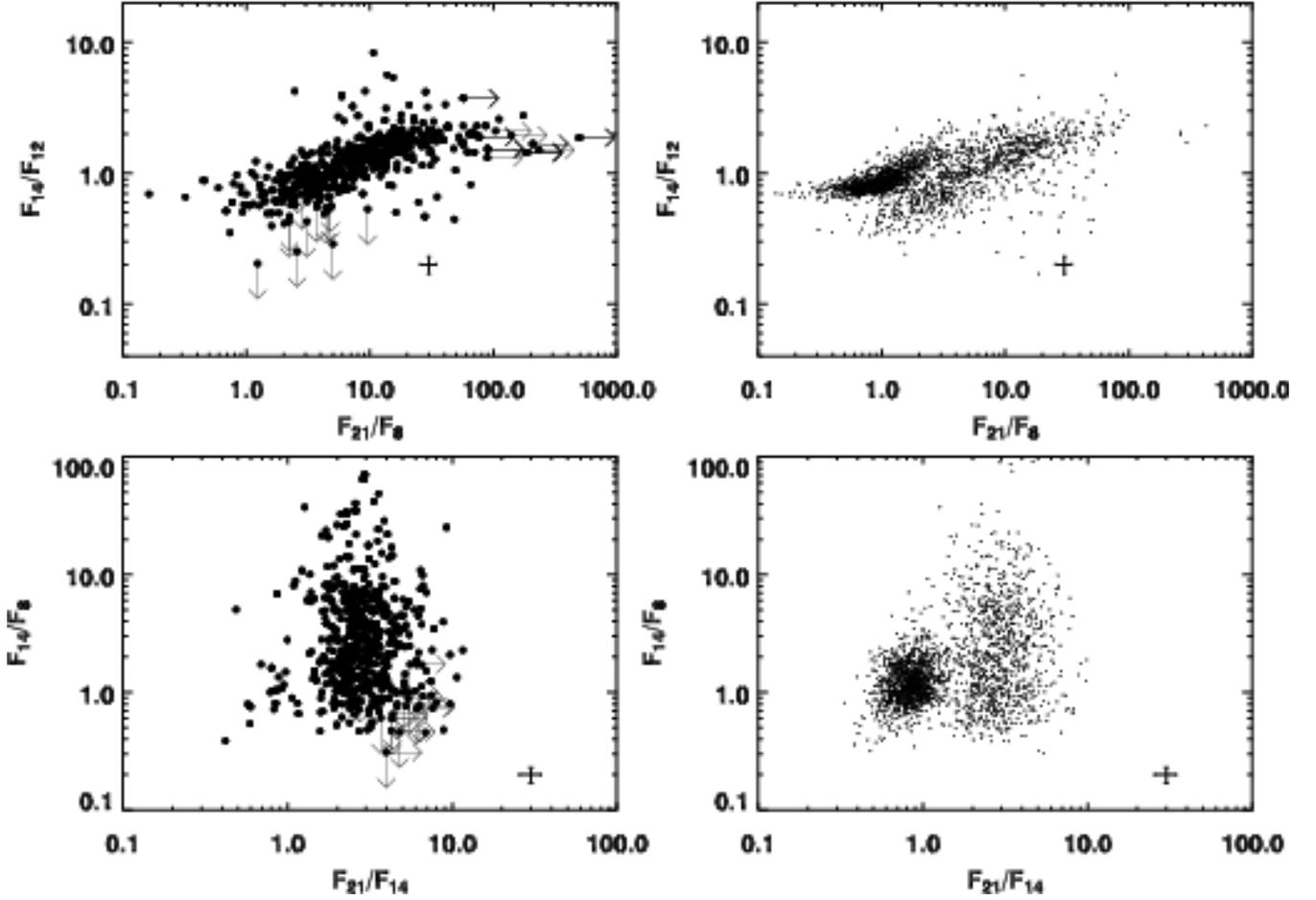}}
\caption{Infrared color-color diagrams for the MSX6C-5 GHz matches (left panels) and for all the high-quality (Lumsden et al. 2002) MSX6C sources in the area of our radio survey (right panels). Representative error bars are plotted. Clearly, radio sources with infrared counterparts belong to the red MSX population.}
\label{radio_cc}
\end{figure}

We now investigate the infrared colors of the 687 radio sources with highly reliable MSX6C counterparts. Lumsden et al. (2002) showed with an earlier version of the MSX Galactic plane catalog (MSX5C) that there are two underlying populations distinguished by their infrared colors. One population shows 'red' infrared colors, and is nebular in nature -- \hii regions, PNe, masers, and young stellar objects. Lumsden et al. (2002) have shown that these objects are heavily embedded sources, in which the infrared colors are determined mainly by the dust surrounding them and not by the central exciting star. The other population shows bluer colors, and consists mainly of evolved stars. Lumsden et al. (2002) based this result on detections of previously classified sources. In the left panels of Figure \ref{radio_cc} we plot two color-color diagrams for our sample of the MSX6C-5 GHz matches following Lumsden et al. (2002) For comparison, we plot
in the right panels all the MSX6C sources with high-quality flags detected in the corresponding bands in the area covered by the radio survey (quality flag criterion as in Lumsden et al. 2002). This plot is slightly different from the parallel plot in Lumsden et al. (2002) which includes only sources in a narrow longitude range ($20^o\le l\le 30^o$), and in a broader latitude range ($|b|\le 6^o$).

Matching the 5 GHz and MSX6C catalogs filters out the majority blue population, leaving mostly red-population sources. For example, there are very few MSX6C-5 GHz matches with $F_{21}/F_8 <2$ and $F_{14}/F_{12} >0.7$ (upper panels) or with $F_{21}/F_{14} <2$ (lower panels). This result implies that our subset of matching sources consists mostly of \hii regions and PNe, although we cannot distinguish between these two populations. From our subset of 687 sources, 29 were identified as PNe by Becker et al. (1994), by matching the older version of the 5 GHz maps and the IRAS catalog, and using the IRAS color criteria. From these 29, 19 were ``candidates'' because only limits on their infrared flux densities were available. By considering the increase in source detection in the present version of the 5 GHz maps, we estimate that there are $\ltorder75$ PNe in our sample ($\ltorder11\%$).
In the following sections we show results that support the argument that the matching subset is dominated by \hii regions.

\subsection{Spatial Distributions}
\label{rad_spat}

\begin{figure}
\centerline{\epsfxsize=6.0in\epsfbox{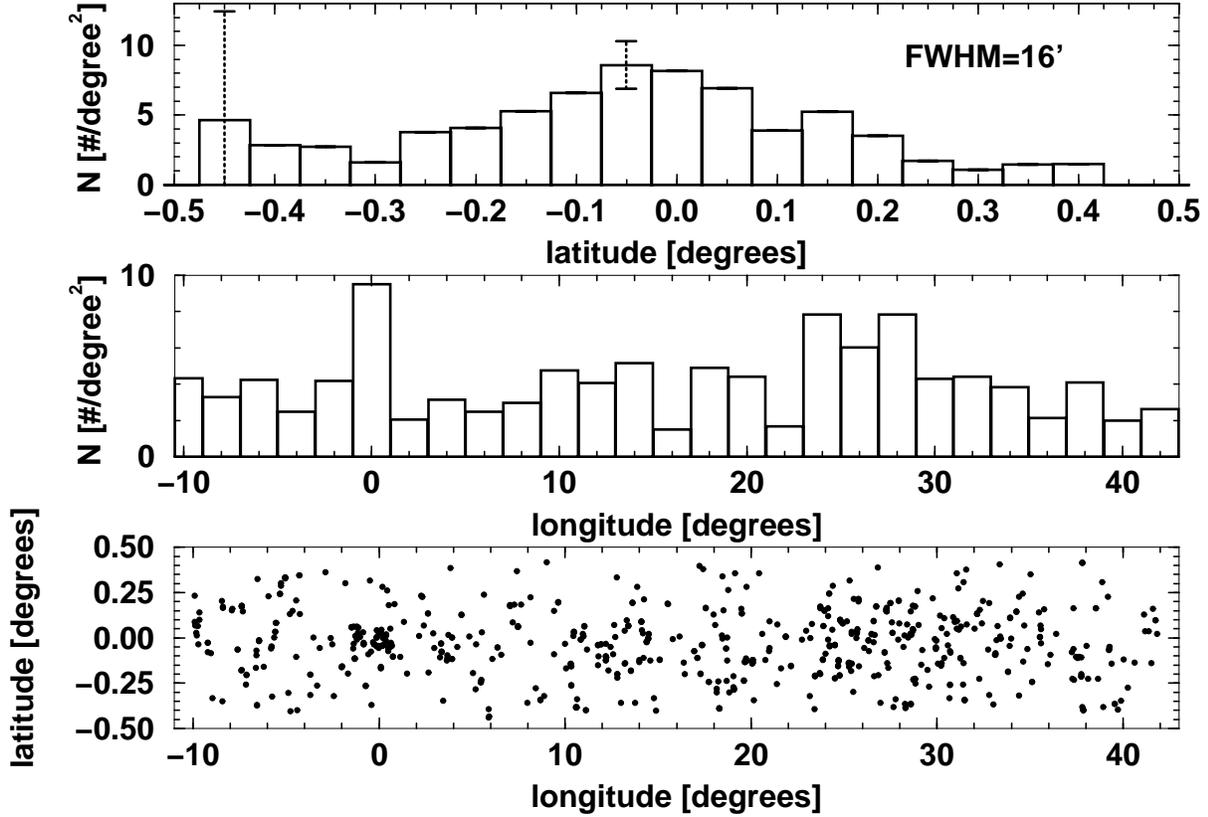}}
\caption{Spatial distributions of the 5 GHz sources with MSX6C matches. The latitude (top) and longitude (middle) distributions are corrected for the uneven coverage. We give the FWHM for the latitude distribution of the matches. The two-dimensional distribution is shown in the bottom panel. The Galactic center contributes significantly to the narrowness of the latitude distribution, but the distribution is quite narrow ($25'$ for $l>10^o$) even if it is excluded. The error bars in the top panel reflect the difference in the number of sources and in the area covered between the central and outer bins. Maxima in longitude are seen at the Galactic Center and at the tangent to the Sagittarius spiral arm ($25<l<30$).}
\label{coords_6cm}
\end{figure}

In Figure \ref{coords_6cm} we show the distribution in Galactic coordinates of the 687 radio sources with MSX6C counterparts. The latitude and longitude histograms are normalized using our survey coverage map (Becker et al. 2004) to correct for uneven coverage due to varying rms levels. The coverage was added up as a function of flux density, and each flux bin was assigned the corresponding coverage map (assuming that the flux distribution is the same over the entire area covered). The raw latitude and longitude bins were then divided by the corresponding coverage area bin, with the underlying assumption that latitude and longitude are independent of each other in each bin.

The 5 GHz sources detected in the infrared show a latitude distribution narrower than the radio catalog in general, with a FWHM of only $16'$ (cf. Becker et al. 1994). For a distance of 8.5 kpc to the Galactic center, this corresponds to $\sim 40$ pc. The narrowness of this distribution suggests that the distribution is dominated by Population I sources. Since these sources are also shown to be mostly embedded sources, we argue that they are predominantly young compact \hii regions. There is a latitude-independent component in the distribution, which can be attributed to the PNe population and/or extragalactic contamination. These populations should have a uniform latitude distribution in the survey area. We estimate their contribution to be $\ltorder 20\%$ of the sample - consistent with our estimate from \S \ref{rad_ir}.
The mean of the latitude distribution of the matching subset is $-0.03^o\pm 0.02^o$. The latitude distribution varies as a function of longitude both in its mean and its width. Sources at $350^o\le l<0^o$, $0^o\le l<10^o$, and $10^o\le l\le 42^o$ have latitude means of $0.00^o\pm 0.02^o$, $-0.01^o\pm 0.02^o$, and $-0.05^o\pm 0.02^o$, respectively. These values are consistent with the known Galactic plane tilt with respect to $b=0^o$ (e.g., Hammersley et al. 1995). Their FWHM is $10'$, $10'$, and $25'$, respectively - narrowing significantly towards the Galactic center.

\subsection{Radio Spectral Indices}
\label{index}

Another useful tool for identifying populations is the shape of the radio spectrum. We calculate the spectral indices by employing a revised 1.4 GHz catalog based on the same data presented by Zoonematkermani et al. (1990). As with the 5 GHz data, the 1.4 GHz data were reanalyzed with improved data reduction techniques (Becker et al. 2004). The radio emission from thermal sources comes from free-free processes and typically has a flat or inverted spectral slope $\alpha\ge -0.1$ ($F_{\nu}\propto\nu^{\alpha}$) in this frequency range. Becker et al. (1994) demonstrated that sources with flat or inverted radio spectral indices have a very narrow distribution of Galactic latitude. Comparing the two radio catalogs and finding which MSX6C-5 GHz matches have a flat or inverted slope can distinguish thermal sources (e.g., \hii regions and PNe) from other, non-thermal radio sources (e.g., SNRs and extragalactic objects).

\begin{figure}
\centerline{\epsfxsize=5.0in\epsfbox{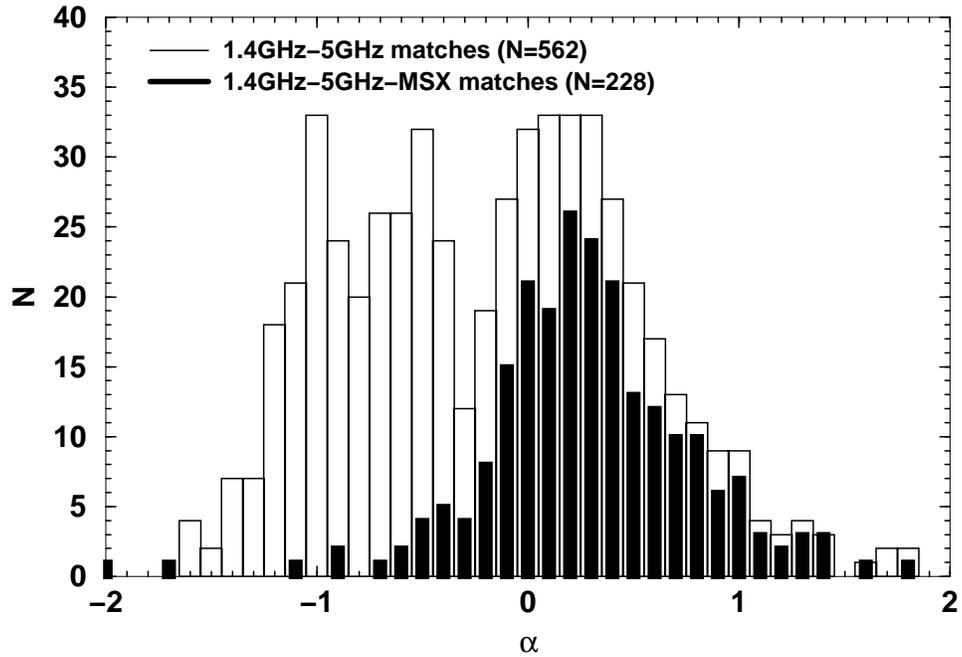}}
\caption{Distributions of spectral slope between 1.4 and 5 GHz using peak flux densities. We show distributions for the whole 5 GHz catalog and for the subset of sources with MSX6C counterparts (shaded histogram). The matching process beautifully filters out the non-inverted sources ($\alpha\ltorder -0.1$).}
\label{alpha_rad}
\end{figure}

Only 562 sources from the 5 GHz catalog have 1.4 GHz counterparts owing primarily to the poorer sensitivity of the latter catalog. Figure \ref{alpha_rad} shows the radio spectral slopes between 1.4 and 5 GHz for the whole 5 GHz catalog and for the subset of sources with MSX6C counterparts (228 sources). The spectral indices shown were calculated using peak flux densities; using the integrated flux densities produces similar results.
The infrared-radio matching neatly removes steep-spectrum sources from the bimodal slope distribution. The average spectral index for the subset of MSX matches is $0.28\pm 0.03$. While this accurately reflects the statistical behavior of the sample, the individual indices are often not reliable owing to the complexity of the radio source morphologies.

There are 52 sources, detected at both 5 and 1.4 GHz, which have spectral indices $\ge -0.1$ but which do not have MSX6C counterparts. There are an additional 41 sources with $\alpha\ge -0.1$ which are cataloged in the MSX6C but have low reliability ($\le 90\%$) radio matches and are therefore not included in the distribution of MSX6C matches in Figure \ref{alpha_rad}. The 52 sources are spread over the entire coordinate range of the radio survey and have, on average, weaker 5 GHz flux densities (median of $\sim 20$ mJy compared to $\sim50$ mJy), which implies that these sources are probably just too weak to be detected by the MSX. 
The latitude distribution of these sources has a component that peaks at $b=0^o$, but its relative fraction is smaller than the MSX6C-5 GHz matching subset, suggesting that most of these sources represent an extragalactic population.

Since most of the sources in our sample are previously unclassified, spectral observations are needed in order to confirm their source type: molecular or recombination lines will allow distance determination and thus the sources' luminosities; infrared spectral shapes can distinguish between \hii regions, PNe and extragalactic sources. If none of the missing 52 sources were extragalactic, then the MSX6C would have been 80\% complete in detecting thermal sources ($1-{{52}\over{228+52}}$), reducing the total completeness level for embedded O star detections to 75\% for optically thick nebulae. Since at least some of them are extragalactic, these completeness levels are lower limits. The completeness level for optically thin nebulae depends only on the MSX completeness and is thus $\ge 80\%$.
The Spitzer legacy program for mapping the Galactic plane (GLIMPSE) will have improved sensitivity and angular resolution, and will be able to produce an even more complete sample of UC \hii regions.

\subsection{Morphology}
\label{morphology}

\begin{figure}
\centerline{\epsfxsize=5.0in\epsfbox{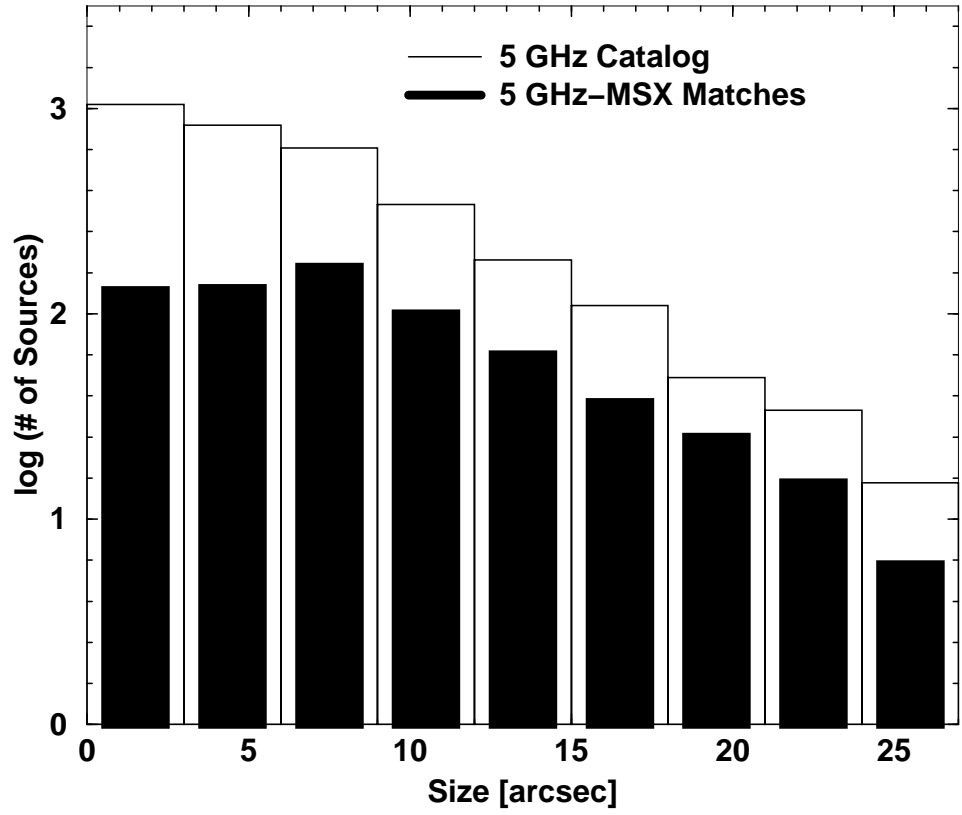}}
\caption{Histogram of 5 GHz sizes for the entire radio catalog (empty bars), and for the MSX6C-5 GHz matching subset (filled bars).}
\label{sizes}
\end{figure}

We use the average of the deconvolved major and minor axes of the 5 GHz sources as a measure of their sizes. The synthesized beam dimensions typically are $9''\times 4''$, with the $9''$ running north-south. We compare the size distribution of the 5 GHz sources with MSX6C matches to the distribution of non-MSX6C sources. On average, the infrared-detected sources have larger sizes: $8.5''\pm 0.2''$ (median of $7.4''$) compared to $6.2''\pm 0.1''$ (median of $4.8''$). The two subsets also show different average major axis to minor axis ratios: $1.70\pm 0.05$ (median of 1) and $2.15\pm 0.04$ (median of 2), respectively. However, the non-MSX6C subset is less than half as bright at 5 GHz on average: $29\pm 3$ mJy (median of 10 mJy) and $12.3\pm 0.02$ mJy (median of 3 mJy), respectively.
This likely arises from a larger extragalactic component for the non-MSX subset which we would expect to be dominated by faint point sources along with some radio doubles (elongated sources with high major-to-minor-axis ratios) at the brighter end. This could also be because many radio sources are in fact side-lobes. While the match with the MSX6C catalog filters out these sources, the non-MSX subset may suffer from this effect.
In Figure \ref{sizes} we show a histograms of the size distribution of our 5 GHz sources and of the subset of MSX6C-5 GHz matches.
The first bin (size $\le 3''$) consists of unresolved sources. This plot shows clearly that extended 5 GHz sources are more likely to have MSX6C counterparts.

\section{Conclusions}
\label{conc}

We have presented a new catalog of UC \hii regions in the first Galactic quadrant. This is the most sensitive, high-resolution catalog of its kind to date, with a completeness for detecting embedded O stars at the $\ge 75\%$ level, depending on the optical depth of the nebula to free-free emission. We have investigated the infrared colors, the spatial distributions, the radio spectral slopes, and the morphologies of sources in this catalog. Our main findings are:
\begin{itemize}
\item[1.] The subset of 5 GHz sources with MSX6C colors is tightly confined to the Galactic plane (FWHM of $16'$ or $\sim 40$ pc). This suggests that the sample is dominated by Population I objects, although our 5 GHz radio catalog and MSX6C infrared data could not distinguish \hii regions from PNe.
\item[2.] By matching our 5 GHz radio catalog and the MSX6C catalog, most of the blue MSX population is filtered out; the majority of the matching subset belongs to the red MSX population. 
\item[3.] By matching our 5 GHz radio catalog and the MSX6C catalog, the non-thermal radio sources are filtered out. The initial bimodal distribution of the radio spectral indices between 1.4 and 5 GHz becomes a single-peaked distribution with $<\alpha>=0.28\pm 0.03$.
\item[4.] The MSX6C does not detect all of the Galactic plane thermal sources; relatively weak 5 GHz sources with flat or inverted radio spectra are not detected by MSX, probably because they are below the MSX detection limit. The latitude distribution of these sources suggests a larger extragalactic fraction compared to the subset of MSX matches. We estimate the MSX6C catalog to be $\ge 80\%$ complete in detecting thermal sources.

\end{itemize}
The overall picture emerging from these results is that objects of nebular nature, which are detected in the radio, have red infrared colors, flat or inverted radio spectra, and a spatial distribution that is tightly confined to the Galactic plane. These properties suggest that these sources are dominated by young compact \hii regions, most of them previously unclassified, although PNe and extragalactic sources also make a small contribution to the sample. A study of radio spectral lines could be used to confirm the source classifications and determine distances to produce a 3-dimensional spatial distribution for these objects. Infrared spectroscopy could be used to determine other properties of these sources, such as the shape of the ionizing spectrum, the type of stellar populations, and nebular properties, such as ionization state, gas temperature and density, and metallicity.

The comprehensive compilation from the literature by Paladini et al. (2003) lists only 355 known \hii regions in our survey area, only 44 of which are smaller than $1'$ in diameter (at 2.7 GHz); thus, our new sample represents a 15-fold increase in the number of compact and UC \hii regions in this region.

\section*{Acknowledgments}
This work was supported in part (R.H.B.) under the auspices of the US Department of Energy by Lawrence Livermore National Laboratory under contract W-7405-ENG-48, and the National Science Foundation under grant AST 0206309.
D.J.H. acknowledges the support of the National Science Foundation under grant AST 02-6-55.

\end{document}